\documentclass[letterpaper, 10 pt, conference]{ieeeconf}  

\IEEEoverridecommandlockouts                              

\usepackage{graphicx} 
\usepackage{balance}
\usepackage{siunitx}
\usepackage{quotmark}

\usepackage{array}
\usepackage{url}

\title{\LARGE \bf
Feasibility Analysis of Fifth-generation (5G) Mobile Networks for Transmission of Medical Imaging Data
}

\author{Nicolai Spicher$^{1,\dagger}$, {\em Member, IEEE}, Michael Schweins$^{2,\dagger}$, Lennart Thielecke$^{2}$,\\ Thomas Kürner$^{2}$, {\em Fellow, IEEE}, and Thomas M. Deserno$^{1}$, {\em Senior Member, IEEE} 
\thanks{$\dagger$ Same contribution}%
\thanks{*This work received funding by the German Federal Ministry of Transport and Digital Infrastructure (BMVI) under grant \# VB5GFWOTUB.}%
\thanks{$^{1}$Peter L. Reichertz Institute for Medical Informatics of TU Braunschweig and Hannover Medical School,
        38106 Braunschweig, Germany.}%
\thanks{$^{2}$Institute for Communications Technology,  Technische Universität Braunschweig,
        38106 Braunschweig, Germany.}%
}

\begin{document}

\maketitle
\thispagestyle{empty}
\pagestyle{empty}

\begin{abstract}
Next to higher data rates and lower latency, the upcoming fifth-generation mobile network standard will introduce a new service ecosystem. Concepts such as multi-access edge computing or network slicing will enable tailoring service level requirements to specific use-cases. In medical imaging, researchers and clinicians are currently working towards higher portability of scanners. This includes i) small scanners to be wheeled inside the hospital to the bedside and ii) conventional scanners provided via trucks to remote areas. Both use-cases introduce the need for mobile networks adhering to high safety standards and providing high data rates. These requirements could be met by fifth-generation mobile networks. In this work, we analyze the feasibility of transferring medical imaging data using the current state of development of fifth-generation mobile networks (3GPP~Release~15). We demonstrate the potential of reaching \SI{100}{\mega\bit\per\second} upload rates using already available consumer-grade hardware. Furthermore, we show an effective average data throughput of \SI{50}{\mega\bit\per\second} when transferring medical images using out-of-the-box open-source software based on the Digital Imaging and Communications in Medicine (DICOM) standard. During transmissions, we sample the radio frequency bands to analyse the characteristics of the mobile radio network. Additionally, we discuss the potential of new features such as network slicing that will be introduced in forthcoming releases.  
\end{abstract}

\section{INTRODUCTION}
Recently, there is a trend towards higher portability of medical imaging technology. On the one hand, there are mobile computed tomography (CT) and magnetic resonance imaging (MRI) scanners available that can be brought to the patient's bedside \cite{Wald_2019}. This reduces the risk of transporting critically ill patients and may speed up clinical workflows. Furthermore, mobile trailer units featuring CT or MRI scanners are increasingly available for bringing this technology to remote areas, emergency care \cite{John_2015}, disaster management \cite{Rutty_2007}, or long-time field studies \cite{Sch_tz_2012}. 

In all use-cases, the acquired imaging data has to be transmitted to a central picture archiving and communication system (PACS) server in compliance with data protection standards. Furthermore, time is an important factor for critically ill patients depending on fast diagnosis. 
However, inside the hospital, wireless computer networks based on IEEE 802.11 standards are potential bottlenecks as they are continuously hampered by an increasing number of clients, e.g., smartphones of patients \cite{Mucchi_2020}.
Outside the hospital, mobile networks such as third- (Universal Mobile Telecommunications System (UMTS)) or fourth-generation mobile networks (Long-Term Evolution (LTE)) are limited with respect to available bandwidth and coverage.

The fifth-generation (5G) of mobile communication systems enables -- next to higher data throughput up to 
\SI{20}{\giga\bit\per\second} -- ultra-reliable and low-latency communications and a huge amount of possible subscribers within a network cell \cite{itu-r_m2410-0_2017}. These characteristics are seen as catalysts for introducing mobile communication into different industries, such as  transport, manufacturing, or media. The coverage of 5G networks is expected to be the fastest deployed mobile communication technology in history and is forecast to cover about 60 percent of the world's population by 2026 \cite{ericsson_2020}.

To date, all 5G networks deployed in Germany are in non-standalone (NSA) mode. Therefore, the 5G cells are paired with existing LTE infrastructure which provides the control functions while the data transmission is performed mainly in the 5G frequency bands. 
Hence, in this work, we analyze the feasibility of NSA 5G according to the 3GPP Release 15 for the transfer of medical imaging data. We aim for gaining first-hand insight into the possibilities of the new mobile network standard and discuss the potential of forthcoming releases for the medical domain.

\section{MATERIAL AND METHODS}
\begin{figure*}[t]
    \centering
       \includegraphics[width=.99\linewidth]{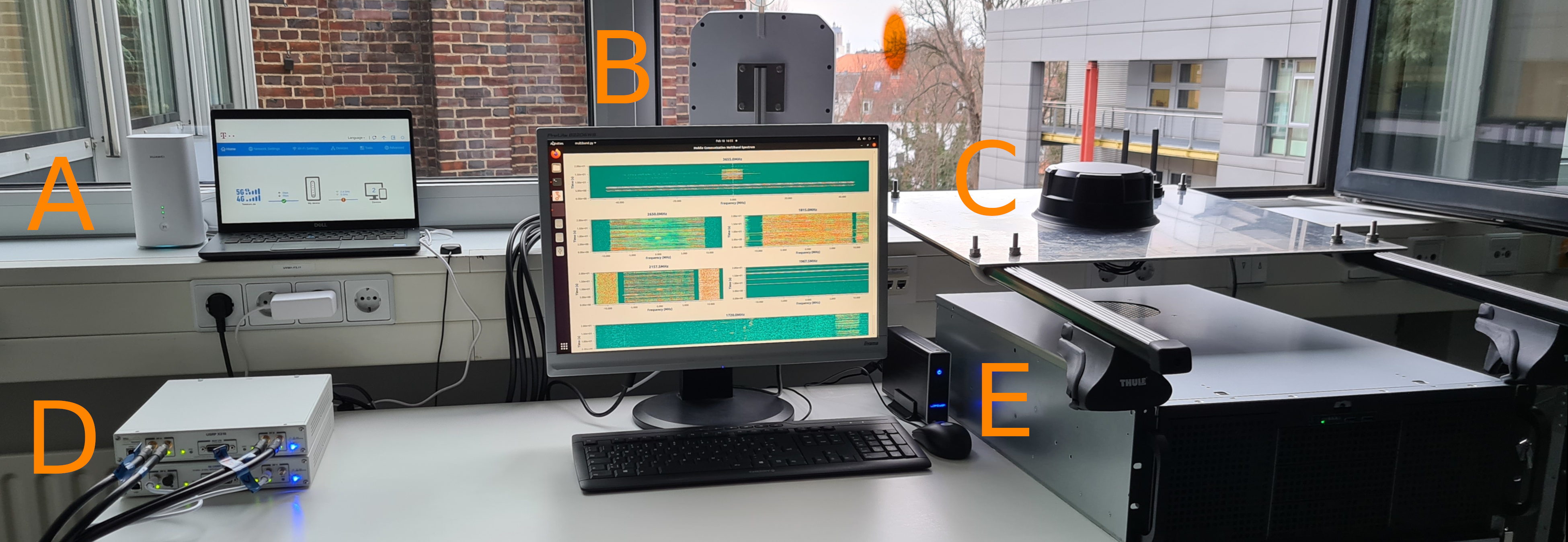} 
            \caption{Experimental setup with direct line of sight to the mobile tower indicated by the ellipsis. A) Laptop serving as DICOM client connected to 5G antenna B) directional antenna C) omni-directional antenna D) USRP devices E) Host-PC. During experiments, the client setup (A) was placed several meters away from the remaining setup to avoid interference and windows were opened to simulate a medical scanner being provided on a mobile trailer.}
            \label{setup}
\end{figure*}

Experiments were performed in the campus area of Brunswick  (approx. 250,000 inhabitants) with direct line of sight from the institute's laboratory to the mobile tower (distance \SI{510}{\meter}). The experimental setup is composed of an off-the-shelf 5G router acting as mobile network client and measurement equipment (Fig. \ref{setup}).

\subsection{Medical Image Data Transmission}
We simulate the use-case that an external medical imaging scanner is outside the local network of a PACS server and transmits imaging data in Digital Imaging and Communications in Medicine (DICOM) format using the 5G mobile network. As PACS we apply the open-source DICOM server \textit{Orthanc} (v1.8.1; \cite{Jodogne_2018}) with the out-of-the-box configuration on an off-the-shelf server (Ubuntu Linux, Intel Core i5, 32GB RAM, 2TB SSD) connected to internet via the network of the Technische Universität Braunschweig.

An off-the-shelf business laptop (Intel Core i7, 16GB RAM, 1TB SSD;  Fig. \ref{setup}:A) acts as DICOM client using the open-source software library \textit{DCMTK}\footnote{https://dicom.offis.de/dcmtk.php.en}. We connect the laptop using Ethernet cable to a HUAWEI 5G~CPE~Pro~2 (Balong 5000 chipset; theoretical 5G transmission rates: \SI{3.6}{\giga\bit\per\second} (downstream), \SI{250}{\mega\bit\per\second} (upstream); Huawei Technologies Co, Shenzhen, China;  Fig. \ref{setup}:A) which provides 5G access via an off-the-shelf internet data plan for business customers with unlimited volume (\textit{Business Mobil XL Plus}, Deutsche Telekom AG, Bonn, Germany). 

The DICOM dataset consists in total of \SI{10}{\giga\byte} and was acquired during an ultra-high-field MRI phantom study (\textit{Magnetom Terra}, Siemens Healthineers, Erlangen, Germany) imaging a fluid-filled phantom. We send the data consecutively in batches of \SI{1}{\giga\byte}. 
Additionally, to gain insight into the maximum data upload rate, we conduct a speed test\footnote{https://www.speedtest.net/}. This allows for estimating the data throughput without any reduction due to the internet link of the DICOM server or the DICOM protocol.

\subsection{Mobile Network Analysis}
During the transmission of the DICOM dataset as well as the speed test we captured radio samples of the mobile radio frequency bands which are licensed by Deutsche Telekom AG. As radio frontend devices we used USRP-2954R and B210 (National Instruments, Texas, Unites States; Fig. \ref{setup}:D) as well as the USRP-X310 (Ettus, Texas, United States; Fig. \ref{setup}:D). The captured samples from the software defined radios are transmitted to a high-performance desktop PC (Fig. \ref{setup}:E) via Ethernet and processed with \textit{GnuRadio}\footnote{https://www.gnuradio.org/} (v3.8).

Using this setup, we are able to receive 4G and 5G signals on different frequency layers (Table \ref{tab:FreqBands}). Both transmission technologies 4G and 5G are based on Orthogonal Frequency Division Multiplex (OFDM). It enables splitting a frequency band into multiple carriers and the time into fixed slot duration, which constitutes single resource elements. Those elements can individually be utilized for uplink (UL) or downlink (DL) transmissions. Depending on the available spectrum in a frequency band, UL and DL transmissions are separated in time (Time Division Duplex, (TDD)) or frequency (Frequency Division Duplex, (FDD)). The n1 band is operated in Dynamic Spectrum Sharing (DSS) mode which allows for using it in parallel for 4G and 5G transmission modes which has been rolled-out recently by the Deutsche Telekom AG.

In this work, we limit our analyses of the mobile network to the energy spread within the mobile radio frequency bands.

\section{RESULTS}

\subsection{Medical Image Data Transmission}
DICOM data was transmitted in all experiments without any data corruption or loss. Reading the webinterface of the 5G router showed peak upload rates of \SI{90}{\mega\bit\per\second}. We stored the duration it took to transfer each batch of \SI{1}{\giga\byte} data. Computing mean and standard deviation results in \SI[parse-numbers=false]{144.91\pm 9.34}{\second}. Hence, the average DICOM data throughput using the proposed setup was approximately \SI{55}{\mega\bit\per\second}. 

\begin{table}[t] 
\centering
\setlength\extrarowheight{1pt}
\caption{Frequency bands measured during mobile network analysis. Bands used during data transmission are written in bold.} 
\label{tab:FreqBands}
\resizebox{\columnwidth}{!}{%
\begin{tabular}{c|c|c|c|c}

	\hline
	Gen. & Band & Mode & Bandwidth & Center Frequency \\
	& & & [MHz] & [MHz] \\
	\hline
	4G & n20 & FDD & 10 & 816 (DL), 857 (UL) \\
	\hline
	4G &  \textbf{n3} & FDD & 20 & 1815 (DL), 1720 (UL) \\
	\hline
	4G/5G & n1 & FDD & 15 & 2157.5 (DL), 1967.5 (UL) \\
	\hline
    4G & 	n7 & FDD & 20 & 2650 (DL), 2530 (UL) \\
	\hline
	5G & \textbf{n78} & TDD & 90 & 3655 \\
	\hline
\end{tabular}
}
\end{table}

\subsection{Mobile Network Analysis}
\begin{figure*}
    \centering
    \includegraphics[width=.92\linewidth]{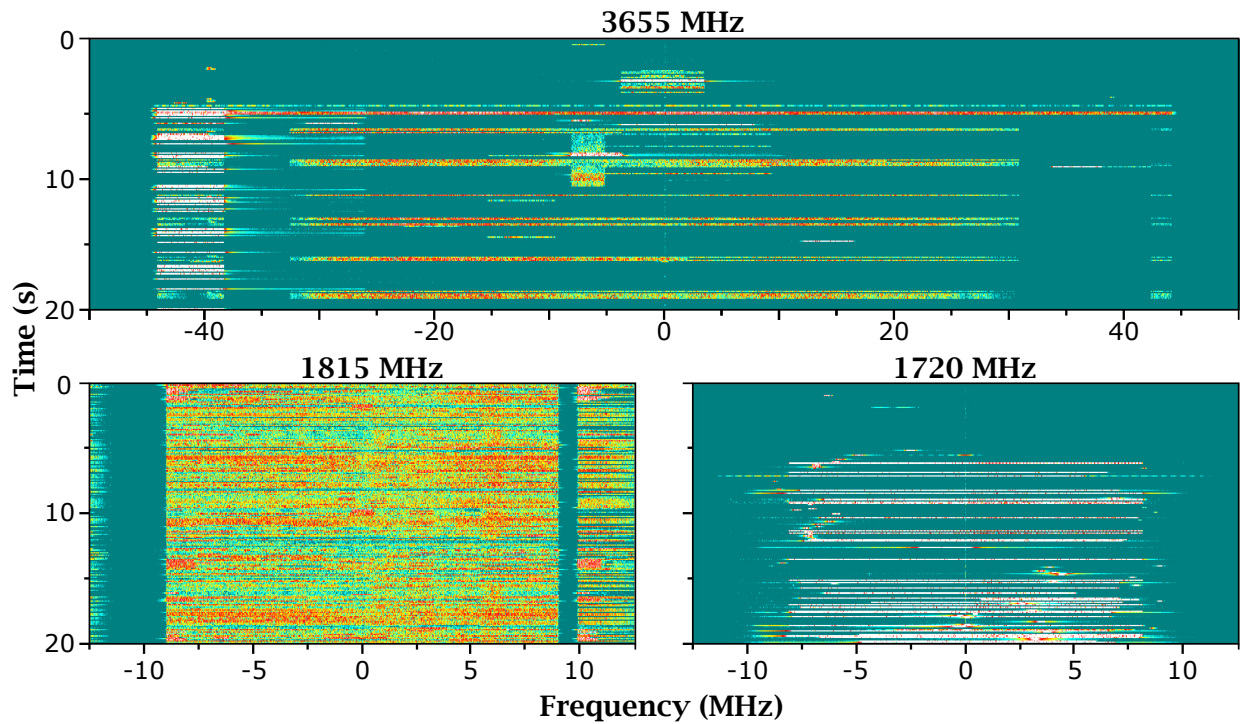}
    \caption{Mobile network analysis during DICOM UL transmission. The transmission starts at approximately \SI{4}{\second}. \textit{Top:} Spectrogram of frequency band n7 (5G). \textit{Bottom:} Spectrogram of frequency band n3 (4G) with the DL being shown on the left (\SI{1815}{\mega\hertz}) and the UL on the right (\SI{1720}{\mega\hertz}). 
    }
    \label{fig:Spec_DICOM}
\end{figure*}

\begin{figure*}
    \centering
    \includegraphics[width=.92\linewidth]{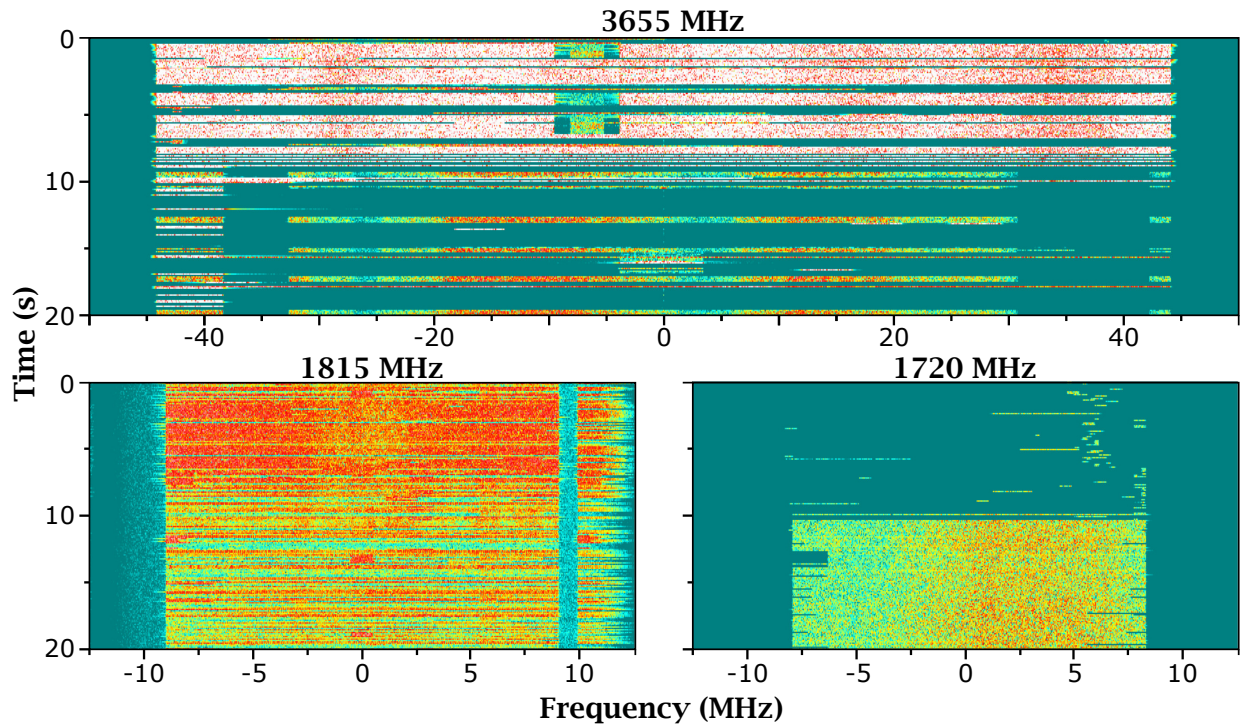}
    \caption{Mobile network analysis during speed test. Until \SI{9}{\second} the DL phase of the speed test is performed, afterwards the UL phase commences.
    \textit{Top:} Spectrogram of frequency band n78 (5G). \textit{Bottom:} Spectrogram of frequency band n3 (4G) with the DL being shown on the left (\SI{1815}{\mega\hertz}) and the UL on the right (\SI{1720}{\mega\hertz}). }
    \label{fig:Spec_Speedtest}
\end{figure*}

We recorded radio frequency samples for all bands shown in Table \ref{tab:FreqBands}. Only bands n3 and n78 were used during transmissions. Spectrograms for the DL and UL channel of band n3 and the channel of band n78 are shown during DICOM transmission (Fig. \ref{fig:Spec_DICOM}) and the speed test (Fig. \ref{fig:Spec_Speedtest}).

Time is depicted on the ordinate, frequency on the abscissa and the amplitude of a particular frequency at a particular instant of time is pseudo-colored. The individual gains of the hardware receiving channels are chosen in a way, that the noise floor is not lifted. The lower end of the colorbar (dark green) is chosen to be just above the noise floor. The colorbar ranges from green over yellow and red to white indicating the highest amplitudes.

The spectrograms depict the baseband signal of the particular frequency band. Although the bandwidth of n78 and n3 is \SI{90}{\mega\hertz} and \SI{20}{\mega\hertz} respectively, only \SI{88}{\mega\hertz} and \SI{18}{\mega\hertz} are utilized to prevent out of band interference. The unused frequency carriers are called guard carriers. The left-bottom graph shows the gap of \SI{1}{\mega\hertz} between the \SI{20}{\mega\hertz} wide 4G band and the adjacent band. In the n78 band no adjacent mobile communication layer could be seen in the top graph. The adjacent lower frequencies are licensed by Telefonica, the upper ones are intended for private licensing in Germany. Since we do not detect any energy there we deduce that there are no devices operating on these bands around the institute.

Regarding the transmission of the DICOM data (Fig. \ref{fig:Spec_DICOM}), data transfer commences at \SI{4}{\second} on the time axis, clearly visible by the white and red areas in the spectrogram of n78 and the uplink band of n3. Since the router is located near the software defined radios we measure a pronounced UL signal.

Regarding the speed test (Fig. \ref{fig:Spec_Speedtest}), it consists of three stages. First, the latency is determined measuring the response time of a dedicated test server. Afterwards, the maximum download data rate is detected while transmitting data from a test server to the device. At around \SI{9}{\second} on the time axis, the test switches to the third stage, in which the device uploads data to the test server to determine the maximum upload data rate. In the band n3 the router is continuously scheduled by the base station to transmit data on nearly all available frequency carriers. In contrast, gaps of approx. \SI{2}{\second} in the frequency range from \SI{-32}{\mega\hertz} up to \SI{31}{\mega\hertz} are clearly visible in the band n78 which is characteristically for the TDD transmission mode. Additional resources at the edges of the frequency band are scheduled to the router, sometimes also within the gaps. The unused resources in between those gaps are dedicated for DL transmission only.

A comparison of Figs. \ref{fig:Spec_DICOM} and \ref{fig:Spec_Speedtest} shows that for the transmission of the DICOM data set the mobile radio network connection is clearly not the limiting factor. In Fig. \ref{fig:Spec_DICOM} a lot of unused UL resources imply that they are not used for uploading DICOM data. 

Additionally, we used the speed test, to evaluate the impact of outdoor-to-indoor attenuation by comparing its results when closing the metal-coated windows. Results show a decrease in maximum data rate for DL and UL transmission. When closing the windows the router reports a \SI{28}{\deci\bel} weaker reference signal leading to a decrease of the maximum uplink data rate from approximately \SI{114}{\mega\bit\per\second} to \SI{20}{\mega\bit\per\second}.

\section{DISCUSSION}
5G mobile networks hold the potential of transforming many industries and sectors including healthcare \cite{Shafi_2017,9354923}. At the moment, the roll-out is still at an early stage only covering NSA 5G (3GPP Release 15). Therefore, we conducted this first analysis on gaining initial insight into possibilities and challenges 5G will bring for medical imaging applications. With portable CT and MRI on the rise, their connectivity could be a critical bottleneck, especially in modalities producing large amounts of data, e.g., ultra-high-field MRI \cite{LADD20181}.

\subsection{Discussion of Results}
In this work, we demonstrate that using off-the-shelf hardware and open-source software allows for transmitting DICOM data with more than \SI{50}{\mega\bit\per\second} in an outdoor setting. These results are in line with results reported by Zhai et al. who sent CT image data from an ambulance in motion to a hospital, reporting on an average uplink of approximately \SI{75}{\mega\bit\per\second} \cite{9354923}.

The theoretical limit of the 5G router is significantly higher (\SI{250}{\mega\bit\per\second}) and using a speed test we demonstrated peak UL rates of \SI{114}{\mega\bit\per\second}. However, it should be considered that we performed indoor experiments with windows opened in a non-laboratory scenario. Furthermore, we used \tqt{out-of-the-box} software configurations only and did not adjust any program parameters. Additionally, the DICOM server was connected to the internet via a connection that was beyond our control and could act as a bottleneck. However, this non-optimized setup already allows a theoretical throughput of  \SI{540}{\giga\byte} per day. Usually, a standard CT/MRI study results in DICOM data smaller than \SI{1}{\giga\byte} but due to high patient throughput and multiple scanners being used in parallel the given limit could be reached by large institutions. In summary, 5G could serve as an alternative or fall-back modality, in case wire-based transmission media are not available, for certain medical institutions.

The comparison of DICOM data transmission and the speed test indicates an over-provisioning of the 5G mobile radio network for the demonstrated setup. However, as 5G is still new on the market and observations of the frequency band also indicate no intensive and regular activity, we assume only very few co-users of the mobile network cell with whom the available resources must be shared. 

\subsection{Discussion of Future 5G Technologies}

In the future, 5G networks will offer the possibility to prioritize individual users with a technology called \tqt{network slicing}. A slice creates a new logical network within the public available mobile radio network for a particular service or for dedicated customers. This approach aims for ensuring appropriate isolation and transmission resources for use-cases with specific requirements \cite{3gpp.23.501}.
Especially for continuous, latency-critical, and small data transmissions this technology enables reliable communication in a calculable duration. However, as these slices are limited in bandwidth, it is rather unlikely that they will be a suitable technology for transmitting large DICOM data.

Another technology introduced in the 5G specifications are \tqt{Quality of service} (QoS) classes linked to an individual connection state \cite{3gpp.23.501}. The \tqt{5G QoS Identifier} (5QI) parameter is defined to be an indicator assigned by the base station to each connected device. The 5QI can be used, e.g., for prioritization in the scheduler or for setting the tolerable transmission error. However, it remains to be shown if the additional costs introduced for the mobile network operators (MNOs) in calculating these 5QI are in a suitable relation to the potential profit. The authors are not aware that these technologies are marketed by MNOs, at least in Germany. 

\subsection{Discussion of Other Aspects Influencing Transmission}

Besides the purely technological advantages of the 5G standard, other factors influence the efficiency of the data transmission in mobile networks. This includes aspects like the proper mounting of antennas or attenuation in indoor environments as we have demonstrated using our experiments with opened and closed windows. These aspects need to be considered when using 5G as part of portable MRI or CT scanners inside the hospital with a large number of potential attenuation factors. Another issue is the 5G coverage and distance to the next mobile tower. In this work, we used a distance of \SI{510}{\meter} which results with line of sight in a signal-to-noise-and-interference ratio (SINR) of \SI{20}{\deci\bel}, which is a reasonable value for mobile communications inside mid- to large cities with a proper mounted antenna.

Another possible application is usage of 5G in ambulance vehicles that provide the data of the patient in real-time to the awaiting emergency department \cite{9354923,cdbme}. In these highly mobile scenarios, the change between different mobile towers and dead zones need to be considered.

\section{CONCLUSION}
Fast and safe transmission of large amounts of medical imaging data are a critical parameter for unfolding the full potential of portable medical imaging in patient care. Research driven by accumulating big data for machine learning \cite{Duncan2020} and novel imaging modalities, such as ultra-high-field MRI, becoming available in clinical practice, will further amplify this requirement. Furthermore, the trend towards portable CT and MRI scanners could be a catalyst for the introduction of mobile networks within healthcare.

In this work, we demonstrate that 5G mobile networks are a feasible infrastructure allowing to achieve a DICOM data throughput of \SI{50}{\mega\bit\per\second} in an \tqt{out-of-the-box} setup. By analyzing the energy spread in mobile radio frequency bands, we demonstrate that the mobile radio network offers even more resources allowing to reach UL data rates larger than \SI{100}{\mega\bit\per\second}. 

Future revisions of the 5G standard will introduce new features such as network slicing or 5QI that might be useful for the transmission of medical imaging data.

\addtolength{\textheight}{-12cm}   



%

\section*{ACKNOWLEDGMENT}
All procedures performed were in accordance with the 1964 Helsinki declaration, as revised in 2000.
The authors are thankful to Stefan Maderwald (Erwin L. Hahn Institute for Magnetic Resonance Imaging, University Duisburg-Essen, Germany) for providing the DICOM data.

\balance

\bibliographystyle{IEEEtran}

\begin{thebibliography}{10}
\providecommand{\url}[1]{#1}
\csname url@samestyle\endcsname
\providecommand{\newblock}{\relax}
\providecommand{\bibinfo}[2]{#2}
\providecommand{\BIBentrySTDinterwordspacing}{\spaceskip=0pt\relax}
\providecommand{\BIBentryALTinterwordstretchfactor}{4}
\providecommand{\BIBentryALTinterwordspacing}{\spaceskip=\fontdimen2\font plus
\BIBentryALTinterwordstretchfactor\fontdimen3\font minus
  \fontdimen4\font\relax}
\providecommand{\BIBforeignlanguage}[2]{{%
\expandafter\ifx\csname l@#1\endcsname\relax
\typeout{** WARNING: IEEEtran.bst: No hyphenation pattern has been}%
\typeout{** loaded for the language `#1'. Using the pattern for}%
\typeout{** the default language instead.}%
\else
\language=\csname l@#1\endcsname
\fi
#2}}
\providecommand{\BIBdecl}{\relax}
\BIBdecl

\bibitem{Wald_2019}
L.~L. Wald, P.~C. McDaniel, T.~Witzel, J.~P. Stockmann, and C.~Z. Cooley,
  ``{Low-cost and portable {MRI}},'' \emph{Journal of Magnetic Resonance
  Imaging}, vol.~52, no.~3, pp. 686--696, oct 2019.

\bibitem{John_2015}
S.~John, S.~Stock, R.~Cerejo, K.~Uchino, S.~Winners, A.~Russman, T.~Masaryk,
  P.~Rasmussen, and M.~S. Hussain, ``{Brain Imaging Using Mobile {CT}: Current
  Status and Future Prospects},'' \emph{Journal of Neuroimaging}, vol.~26,
  no.~1, pp. 5--15, nov 2015.

\bibitem{Rutty_2007}
G.~N. Rutty, C.~E. Robinson, R.~BouHaidar, A.~J. Jeffery, and B.~Morgan, ``{The
  Role of Mobile Computed Tomography in Mass Fatality Incidents},''
  \emph{Journal of Forensic Sciences}, vol.~52, no.~6, pp. 1343--1349, sep
  2007.

\bibitem{Sch_tz_2012}
U.~H. Schütz, A.~Schmidt-Trucksäss, B.~Knechtle, J.~Machann, H.~Wiedelbach,
  M.~Ehrhardt, W.~Freund, S.~Gröninger, H.~Brunner, I.~Schulze, H.-J. Brambs,
  and C.~Billich, ``{The Transeurope Footrace Project: longitudinal data
  acquisition in a cluster randomized mobile {MRI} observational cohort study
  on 44 endurance runners at a 64-stage 4,486km transcontinental
  ultramarathon},'' \emph{{BMC} Medicine}, vol.~10, no.~1, jul 2012.

\bibitem{Mucchi_2020}
L.~{Mucchi}, R.~{Vuohtoniemi}, H.~{Virk}, A.~{Conti}, M.~{Hämäläinen},
  J.~{Iinatti}, and M.~Z. {Win}, ``{Spectrum Occupancy and Interference Model
  Based on Network Experimentation in Hospital},'' \emph{IEEE Transactions on
  Wireless Communications}, vol.~19, no.~9, pp. 5666--5675, 2020.

\bibitem{itu-r_m2410-0_2017}
\BIBentryALTinterwordspacing
{ITU-R}, ``{M.2410-0 {Minimum} requirements related to technical performance
  for {IMT}-2020 radio interface(s)},'' ITU, Electronic Publication, Geneva,
  Tech. Rep., 2017. [Online]. Available:
  \url{https://www.itu.int/pub/R-REP-M.2410-2017}
\BIBentrySTDinterwordspacing

\bibitem{ericsson_2020}
\BIBentryALTinterwordspacing
F.~Jejdling, ``Ericsson {Mobility} {Report},'' Online Publication, Nov. 2020.
  [Online]. Available: \url{www.ericsson.com}
\BIBentrySTDinterwordspacing

\bibitem{Jodogne_2018}
S.~Jodogne, ``{The Orthanc Ecosystem for Medical Imaging},'' \emph{Journal of
  Digital Imaging}, vol.~31, no.~3, pp. 341--352, may 2018.

\bibitem{Shafi_2017}
M.~Shafi, A.~F. Molisch, P.~J. Smith, T.~Haustein, P.~Zhu, P.~D. Silva,
  F.~Tufvesson, A.~Benjebbour, and G.~Wunder, ``{5G: A Tutorial Overview of
  Standards, Trials, Challenges, Deployment, and Practice},'' \emph{{IEEE}
  Journal on Selected Areas in Communications}, vol.~35, no.~6, pp. 1201--1221,
  jun 2017.

\bibitem{9354923}
Y.~{Zhai}, X.~{Xu}, B.~{Chen}, H.~{Lu}, Y.~{Wang}, S.~{Li}, X.~{Shi},
  W.~{Wang}, L.~{Shang}, and J.~{Zhao}, ``{5G-Network-Enabled Smart Ambulance:
  Architecture, Application, and Evaluation},'' \emph{IEEE Network}, vol.~35,
  no.~1, pp. 190--196, 2021.

\bibitem{LADD20181}
M.~E. Ladd, P.~Bachert, M.~Meyerspeer, E.~Moser, A.~M. Nagel, D.~G. Norris,
  S.~Schmitter, O.~Speck, S.~Straub, and M.~Zaiss, ``{Pros and cons of
  ultra-high-field MRI/MRS for human application},'' \emph{Progress in Nuclear
  Magnetic Resonance Spectroscopy}, vol. 109, pp. 1--50, 2018.

\bibitem{3gpp.23.501}
\BIBentryALTinterwordspacing
3GPP, ``{System architecture for the 5G System (5GS) (Release 16)},'' {3rd
  Generation Partnership Project (3GPP)}, Technical Specification (TS) 23.501,
  12 2020, version 16.7.0. [Online]. Available:
  \url{https://portal.3gpp.org/desktopmodules/Specifications/SpecificationDetails.aspx?specificationId=3144}
\BIBentrySTDinterwordspacing

\bibitem{cdbme}
J.~Gaebel, C.~Bockelmann, A.~Dekorsy, R.~Wendlandt, T.~Musiol, J.~Neumann,
  T.~Neumuth, and M.~Rockstroh, ``{Requirements for 5G Integrated Data Transfer
  in German Prehospital Emergency Care},'' \emph{Current Directions in
  Biomedical Engineering}, vol.~6, no.~3, pp. 9--12, 2020.

\bibitem{Duncan2020}
J.~S. {Duncan}, M.~F. {Insana}, and N.~{Ayache}, ``{Biomedical Imaging and
  Analysis in the Age of Big Data and Deep Learning [Scanning the Issue]},''
  \emph{Proceedings of the IEEE}, vol. 108, no.~1, pp. 3--10, 2020.

\end{thebibliography}


\end{document}